\documentclass[11pt]{article}
\usepackage{graphicx}
\usepackage{amsmath}
\begin{document}

\begin{center}
{\Large\bf Partonic State and Single Transverse Spin Asymmetries in
Semi Inclusive DIS} \vskip 10mm
J.P. Ma and H.Z. Sang    \\
{\small {\it  Institute of Theoretical Physics, Academia Sinica,
P.O. Box 2735,
Beijing 100080, China }} \\
\end{center}
\vskip 10mm
\begin{abstract}
Using partonic states we verify the transverse momentum dependent factorization
for single transverse spin asymmetries in semi inclusive DIS at leading but nontrivial
order of $\alpha_s$. The factorization has been previously derived in a formal way
by using diagram expansion at hadron level. We find with our partonic results that
two relevant structure functions satisfy the factorization. Our results also
satisfy the collinear factorization but with the perturbative coefficient
different than that derived formally.
\end{abstract}
\vskip 1cm

\par\noindent
{\bf 1.} Single transverse-spin asymmetries(SSA) have been observed in experiment. An updated
review about the phenomenology of SSA can be found in \cite{Review}.
These asymmetries are expected if scattering amplitudes have
nonzero absorptive parts. SSA in a scattering of elementary particles
is easy to be understood, but SSA in a process involved with hadrons, especially,
with a transversely polarized hadron, is difficult to be predicted,
because we have not enough information about the structure of hadrons. However, the observed
asymmetries provide a new tool to study the structure because they are sensitive
to the correlations of partons inside a hadron and the angular momenta of these partons.
\par
In general it is difficult to predict the size of SSA except
some cases like production of a heavy quark with transverse polarization,
where SSA can be studied by using perturbative QCD directly\cite{TPL}.
In the cases studied in experiment with an initial hadron being transversely polarized,
it is not possible to use perturbative QCD directly. One needs to rely on QCD factorization
in which perturbative- and nonperturbative effects are separated. The nonperturbative effects
are represented by various matrix elements of QCD operators and these matrix elements contain
information about the structure of hadrons.
Two factorization approaches have been proposed. One is by using
transverse-momentum-dependent(TMD) factorization, where one takes transverse momenta of partons in
hadrons into account. Another is the collinear factorization.
It should be noted that in the two approaches the factorization is derived or proposed rather formally in
the sense that one works at hadron level by using the diagram expansion.
The two approaches are applicable in different kinematic regions. The TMD factorization can only be used
for the kinematic region where the observed transverse momentum is much more smaller than some large scales,
while the collinear factorization can be used if the observed transverse momentum is
much more larger than $\Lambda_{QCD}$.
The purpose of our work is to examine the two factorization approaches for SSA in Semi Inclusive DIS(SIDIS)
at parton level, following our work for SSA in Drell-Yan processes\cite{MaSa}.
\par
In the approach of TMD factorization, the nonperturbative effects responsible for SSA
are represented by matrix elements containing $T$-odd- and spin-flip effects. These matrix
elements are the so-called Sivers\cite{Sivers}- and Collins\cite{JC} functions.
So far TMD factorization has been examined carefully only for physical quantities
which do not contain $T$-odd effects\cite{CS,CSS,JMY,JMYG,CAM}.
TMD parton distributions
entering the factorization can be defined with QCD operators
consistently. Intensive efforts has been spent
to study how to consistently define or interpret Sivers function as a parton distribution
which is gauge invariant and contains initial- or final state interactions\cite{JC,SJ1,TMDJi,Mulders97,Boer03}.
Through these studies consistent definitions of these distributions or fragmentation functions which
contain $T$-odd effects can be given. In \cite{CAM} the problem of universality of $T$-odd parton distributions
and $T$-odd fragmentation functions has been studied. The TMD factorization of Sivers effect
has been examined in a spectator model\cite{SJ1}.
SSA has been studied extensively in terms of Sivers functions
\cite{Anselmino,Mulders,DeSanctis,Efremov,BQMa}.
In the approach of collinear factorization, the nonperturbative effects responsible for SSA
are represented  by twist-3 matrix elements\cite{QiuSt, EFTE,KaKo,EKT},
or called ETQS matrix elements. These twist-3 matrix elements contain only the effect
of spin-flip interactions. The nonzero absorptive part or $T$-odd effect
is generated by poles of parton propagators in hard scattering.
Applications of the collinear factorization for SSA can be found in \cite{tw31,tw32}.
It is interesting to note that for SIDIS and Drell-Yan both approaches are applicable
for certain kinematic region and it can be shown that the two approaches are equivalent\cite{JQVY1,JQVY2}.
\par
As mentioned, the above factorized results are derived rather formally in
the sense that one works at hadron level by introducing various parton density
matrices of hadrons in a diagram expansion.
It should be noted that QCD factorizations,  if they are proven, are
general properties of QCD Green functions. It means that the two factorization approaches,
if they hold, they should also hold by replacing hadrons with partons. It is the purpose
of the study presented here to show how SSA in SIDIS can be factorized in the TMD-factorization
approach by replacing hadrons with partons. In our previous work\cite{MaSa} we have shown
that the TMD-factorization for SSA in Drell-Yan processes holds with partonic states
at leading but nontrivial order of $\alpha_s$. With the partonic states one can also derive
a collinear factorization for SSA in Drell-Yan processes but with perturbative coefficients
different than those derived before.
For SIDIS we replace the hadron in the initial- and final state
with a quark $q$. In order to have spin-flip we keep
the quark mass as nonzero and every quantity is calculated at leading power of $m$.
It should be emphasized that
the derived perturbative coefficients in the factorization formulas do not depend on the quark mass $m$.
Two structure functions for SSA in SIDIS are studied in this work.

\par\vskip20pt
\par\noindent
{\bf 2.} We consider the SIDIS:
\begin{equation}
 \ell + h_A(P,s_\perp) \to \ell + \gamma^*(q) + h_A(P,s_\perp) \to
 \ell + X + h_B(P_h).
\end{equation}
We take a coordinate system in which
$h_A$ moves in the $z$-direction and the virtual photon moves in the $-z$-direction.
The initial hadron $h_A$ is transversely polarized.
We will use the  light-cone coordinate system, in which a
vector $a^\mu$ is expressed as $a^\mu = (a^+, a^-, \vec a_\perp) =
((a^0+a^3)/\sqrt{2}, (a^0-a^3)/\sqrt{2}, a^1, a^2)$ and $a_\perp^2
=(a^1)^2+(a^2)^2$. We also introduce two light-cone vectors: $n^\mu =(0,1,0,0)$ and $l^\mu = (1,0,0,0)$.
The relevant kinematic variables can be defined as:
\begin{equation}
x_B = \frac{-q^2}{2P\cdot q}= \frac{Q^2}{2P\cdot q}, \ \ \ \ z_h = \frac{P\cdot P_h}{P\cdot q}.
\end{equation}
The hadronic tensor for the process is defined with the electromagnetic current $J^\mu$ of quarks as
\begin{equation}
    W^{\mu\nu}=\frac{1}{4z_{h}}\sum_{X}\int\frac{d^{4}x}{(2\pi)^{4}}e^{iq\cdot x}
    \langle h_A (P, s_\perp)\vert J^{\mu}(x) \vert X P_{h} \rangle
    \langle XP_{h}|J^{\nu}(0) \vert h_A(P,s_\perp) \rangle.
\end{equation}
We will be interested in the kinematical region of $P^2_{h\perp} \ll Q^2$,
The hadronic tensor in this region  has the following leading-twist structures:
\begin{eqnarray}
    2W^{\mu\nu}&=&-g_{\bot}^{\mu\nu}F^{(1)}_{U}
    +(g_{\bot}^{\mu\nu}-\hat{P}_{h\bot}^{\mu}\hat{P}_{h\bot}^{\nu}-\hat{P}_{h\bot}^{\nu}\hat{P}_{h\bot}^{\mu})F_{U}^{(2)}
\nonumber\\
    &&  -g_{\bot}^{\mu\nu}\epsilon^{\alpha\beta}s_{\bot\alpha}\hat{P}_{h\bot\beta}F_{T}^{(1)}
        + s_{\bot\alpha}\left ( \epsilon_{\bot}^{\alpha\mu}\hat{P}_{h\bot}^{\nu}+
       \epsilon_{\bot}^{\alpha\nu}\hat{P}_{h\bot}^{\mu}
     -g_{\bot}^{\mu\nu}\epsilon^{\alpha\beta}\hat{P}_{h\bot\beta} \right )\left (F_{T}^{(2)}-F_{T}^{(3)}/2 \right )
\nonumber\\
   && +\hat{\vec{P}}_{h\perp}\cdot\vec{s}_{\bot}\hat{ P}_{h\bot\alpha}
   \left ( \epsilon_{\bot}^{\alpha\mu}\hat{P}_{h\bot}^{\nu} + \epsilon_{\bot}^{\alpha\nu}\hat{P}_{h\bot}^{\mu} \right)
     F_{T}^{(3)}
\end{eqnarray}
with the notation $g_\perp^{\mu\nu} =g^{\mu\nu}-n^\mu l^\nu -n^\nu l^\mu$ and
$\epsilon_{\bot}^{\mu\nu}=\epsilon^{\alpha\beta\mu\nu}l_{\alpha}n_{\beta}$.
$\hat {\vec {P}}_{h\perp}$ is the unit vector  $\hat {\vec {P}}_{h\perp} = \vec P_{h\perp}/\vert  \vec P_{h\perp}\vert$.
All structure functions $F$'s  depend on $x_B$, $z_h$, $P^2_{h\bot}$, and $Q^2$.
The differential cross section is given by:
\begin{eqnarray}
    \frac{d\sigma}{dx_{B}dydz_{h} d\phi_s d^{2}\vec{P}_{h\bot}}&=&\frac{2 \alpha_{em}^{2} S}{Q^{4}}
   \left  \{(1-y+\frac{y^{2}}{2})x_{B}\left [F_{U}^{(1)}+\sin(\phi_{h}-\phi_{s})|\vec s_{\bot}|F_{T}^{(1)}\right ]
\right.
\nonumber\\
    && \left. +(1-y)x_{B}\left [-\cos(2\phi_{h})F_{U}^{(2)}+|\vec{s}_{\bot}|\sin(\phi_{h}+\phi_{s})F_{T}^{(2)}
\right.\right.
\nonumber\\
    &&\left. \left. +\frac{1}{2}|\vec{s}_{\bot}|\sin(3\phi_{h}-\phi_{s})F_{T}^{(3)}\right ] \right \}
\end{eqnarray}
where $S$ is the invariant mass of the initial lepton and the initial hadron, $y$ is the fraction of
the lepton energy loss. $\phi_S$ and $\phi_h$ are azimuthal angles for the transverse polarization vector
of the initial hadron and the transverse momentum of the final state hadron, respectively.
A more general decomposition of the above differential cross section can be found in \cite{Gcro}.
By studying the angular dependence of the differential cross section in experiment, one can measure
the different structure functions. In this letter we will focus on $F_T^{(1)}$ and $F_T^{(2)}$.
\par
It has been suggested that the two structure functions can be factorized with various TMD partons distributions
and TMD fragmentation functions. The relevant TMD parton distributions and fragmentation functions
can be defined by introducing a gauge link
along the direction $u^\mu =(u^+,u^-,0,0)$:
\begin{equation}
L_u (\pm \infty, z) = P \exp \left ( -i g_s \int^{\pm \infty}_0  d\lambda
     u\cdot G (\lambda u + z) \right ).
\end{equation}
Two relevant parton distributions
can be defined in the limit $u^+ \ll u^-$
\cite{JC,JMY,Mulders97}:
\begin{eqnarray}
q_\perp (x,k_\perp)
\varepsilon_\perp^{\mu\nu} s_{\perp\mu}  k_{\perp\nu}
  & =& \frac{1}{2}  \int \frac{dz^- d^2 z_\perp}{(2\pi)^3}
                e ^{-i k \cdot z }
                \langle P, \vec  s_\perp \vert
                \bar\psi (z ) L_u ^\dagger (\infty, z) \gamma^+
                L_u (\infty, 0) \psi(0) \vert P,\vec s_\perp \rangle ,
\nonumber\\
  \delta q_T(x,k_\perp) s_\perp^\mu & =& \frac{1}{2}  \int \frac{dz^- d^2 z_\perp}{(2\pi)^3}
                e ^{-i k \cdot z }
                \langle P, \vec  s_\perp \vert
                \bar\psi (z ) L_u ^\dagger (\infty, z) \gamma^+ \gamma^\mu_\perp \gamma_5
                L_u (\infty, 0) \psi(0) \vert P,\vec s_\perp \rangle,
\end{eqnarray}
with $z^\mu =(0,z^-,\vec z_\perp)$. $x$ is defined as $k^+ = xP^+ $. In the above one only takes
the spin-dependent part of the matrix elements into account. $q_\perp (x,k_\perp)$ is the Sivers
function\cite{Sivers}. A relevant TMD parton fragmentation function is defined with the gauge link
$L_v(-\infty, 0)$ along the direction $v^\mu =(v^+,v^-,0,0)$ with $v^- \ll v^+$:
\begin{eqnarray}
\delta \hat q (x,P_{h\perp})
  P_{h\perp}^\mu
   &=& \frac{1}{4x}  \int \frac{dz^+ d^2 z_\perp}{(2\pi)^3}
                \exp\left (-iz^+ P_{h}^- /x \right )
\nonumber\\
       && \cdot         \sum_X \frac{1}{3} {\rm Tr} \left [ \left ( i \gamma^\mu_\perp \gamma^- \right )
                \langle  0 \vert L_v(-\infty,0) \psi (0) \vert P_h X \rangle
                \langle X P_h \vert \bar\psi (z ) L_v ^\dagger (-\infty, z)  \vert 0\rangle \right ],
\end{eqnarray}
with $z^\mu =(z^+,0,\vec z_\perp)$.
The parton carries the momentum $k^\mu =(0,P_h^-/x,0,0)$ and the hadron is with
the momentum $P^\mu_{h} = (P_h^+,P_h^-,{\vec P}_{h\perp})$ with $P_h^-$ as the large component.
The function  $\delta \hat q$ is called as Collins function\cite{JC}.
It should be noted that one can also define the Collins function in a frame through a Lorentz transformation
where the hadron has zero transverse momentum and the parton has the transverse momentum $-x {\vec P}_{h\perp}$
\cite{CS}.
The Sivers- and Collins function
can only be nonzero if a nonzero absorptive part
exists in the corresponding hadronic matrix elements. The definitions of other TMD parton distributions
or fragmentation function of
a unpolarized hadron  can be found in \cite{CS,JMY,JMYG}.
One should keep in mind that all of these TMD parton distributions and fragmentation functions
are defined in non-singular gauge. In principle in these definitions
one should add a gauge link in the transverse direction at the space-time boundary as shown in \cite{TMDJi},
e.g., in Eq.(8) one should add a gauge link at $x^- =\infty$ along the transverse direction to make
the definition gauge-invariant explicitly.
However, in a non-singular gauge like Feynman gauge the gauge fields
at infinite $x^+$ or $x^-$ are zero. Hence the gauge links at infinite $x^+$ or $x^-$ are unit matrices.
In a singular gauge like light-cone gauges these gauge links can not be omitted\cite{TMDJi}.
We will work with Feynman gauge.
The structure functions
can be factorized as:
\begin{eqnarray}
F_T^{(1)}(x_B, z_h, P_{h\perp}) &=&  \int d^2 k_\perp d^2 p_\perp \left ( \vec k_\perp \cdot\hat{\vec{P}}_{h\perp} \right ) q_\perp (x_B,k_\perp)
     \hat q(z_h, p_\perp ) \delta^2 (z_h \vec k_\perp + \vec p_\perp -\vec{P}_{h\perp} ),
\nonumber\\
F_T^{(2)}(x_B, z_h, P_{h\perp}) &=& \int d^2 k_\perp d^2 p_\perp  \left ( \vec k_\perp \cdot\hat{\vec{P}}_{h\perp} \right ) \delta q_T (x_B,k_\perp)
     \delta \hat q(z_h, p_\perp ) \delta^2 (z_h \vec k_\perp + \vec p_\perp -\vec{P}_{h\perp} ),
\end{eqnarray}
where $\hat q$ is the standard TMD parton fragmentation.
Beyond the leading order of $\alpha_s$
one has to implement a soft factor representing effects of soft-gluon radiation.
\par
For our purpose we replace each hadron state with a one-quark state. We consider the SIDIS of the quark states as:
\begin{equation}
 q(P,s_\perp) + \gamma^* (q) \to q(P_h) +X,
\end{equation}
with $P^+$ and $P_h^-$ as large momentum components. In order to generate SSA  and nonzero
$q_\perp$, there must be exchange of two gluons at the leading order. One of the two gluon
must be in the unobserved state $X$ because $P_h$ has nonzero transverse components.
 The exchange of another gluon must introduce a virtual contribution and
generate a nonzero absorptive part in the scattering amplitude. With this requirement one easily
finds nonzero contributions to $F_T^{(1,2)}$ come from the interference of amplitudes
given by diagrams in Fig.1.
\par
\begin{figure}[hbt]
\begin{center}
\includegraphics[width=14cm]{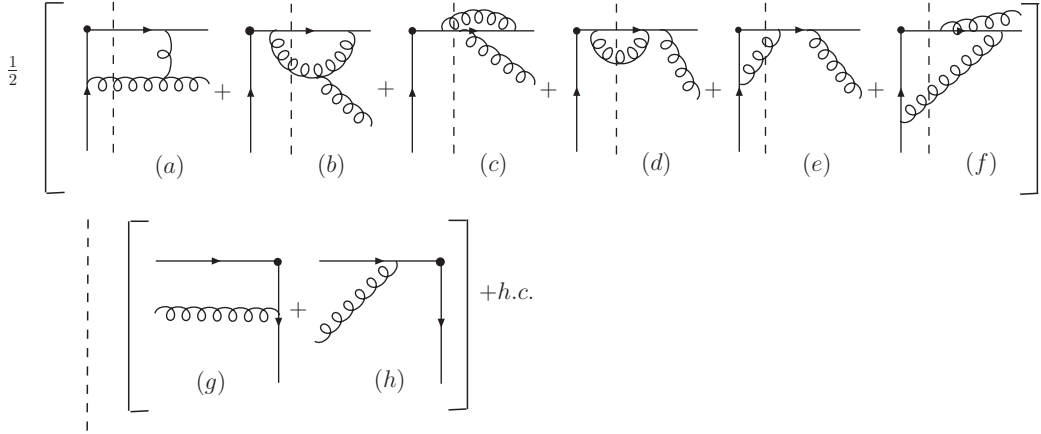}
\end{center}
\caption{Diagrams for the contributions to the relevant structure functions. The broken lines
are cuts. Black dots represent the insertion of the electromagnetic current.  }
\label{Feynman-dg1}
\end{figure}
\par
In Fig.1 the diagrams in the first raw are the scattering amplitudes which can have a nonzero
absorptive part. The absorptive parts are determined by physical cuts which are given
in these diagrams. To calculate the relevant hadronic tensor in the full kinematic region
one has to calculate all diagrams in Fig.1. This will be very tedious. Fortunately we only need
the leading contribution in the expansion of $P^2_{h\perp}/Q^2$. As shown in \cite{MaSa},
one can perform the expansion before the loop-momentum integration of the two exchanged gluons.
Before presenting our detailed results we point here that
with our partonic state $F_T^{(3)}$ does not receive any nonzero contribution
at the orders we consider. It is worth to point out that
the parton distribution function related to $F_T^{(3)}$
is also zero in the quark target model of \cite{H1T}.
 We also find that the contributions to $F_T^{(1,2)}$
from gluon fragmentation are power suppressed by $P^2_{h\perp}/Q^2$ or by $\alpha_s$.
This fact is in agreement
with the factorization formulas in Eq.(9).
In the below we will in turn show the factorization for $F_T^{(1)}$ and $F_T^{(2)}$.
\par\vskip20pt
\noindent
{\bf 3.} We first study the factorization of $F_T^{(1)}$.
As mentioned in the above, the full calculation is very tedious. Instead of doing this
we expand contributions with loop integrals in $q^2_\perp /Q^2$ first and then perform the loop integral.
A convenient way for the expansion is to analysis different regions of
the loop momentum. The results should be the same as those obtained from
the full calculation by taking the limit $q^2_\perp \ll Q^2$.
If one wants to have results of the whole $q^2_\perp$-region to test the factorization,
one has to do the full calculation.
\par
We denote the momentum of the gluon in the intermediate state as $k_g$  and the momentum
of the virtual gluon as $k$. Each contribution from Fig.1 can be written as:
\begin{equation}
\int \frac{d^4 k}{(2\pi)^4} \frac{ d^4 k_g} {(2\pi)^4} (2\pi) \delta (k_g^2)
               (2\pi)^4 \delta^4(P +q -P_h -k_g )
     \frac{1}{D_1 D_2 D_3 D_4 D_5 } \cdot {\rm Tr} \left [ \cdots \right ],
\end{equation}
In each contribution there are
five propagators, their denominators are denoted as $D_i(i=1,2,\cdots, 5)$.
The numerator represented in the above as ${\rm Tr} \left [ \cdots \right ]$
is a trace of product of $\gamma$-matrices.
We scale the momentum $\vec k_{g\perp} =-\vec P_{h\perp}$ as at order of $\lambda$.
and expand each contribution in $\lambda$.
Since the transverse momenta are at order of $\lambda$, the leading order
contributions comes from the integration region of $k$ with $k_\perp$ at order
of $\lambda$ or smaller, i.e., from the region with ${\vec k}_\perp \sim (\lambda, \lambda)$
or ${\vec k}_\perp \sim (\lambda^2, \lambda^2)$.
Without performing the integration of $k$ we can determine the order of $\lambda$ of each contribution.
We use $d$ to denote the leading power of a contribution by setting its numerator as $1$
and $n$ as the leading power of its numerator. Hence the leading order of the contribution
is at $\lambda^{d+n}$. It seems that without a detailed calculation of numerators
it is unclear how to determine $n$. But we know $n\geq 1$. It is rather easy to determine $d$
of each contribution by power counting. We find that only those diagrams have
the leading power with $d=-4$. They are the interferences of Fig.1a to Fig.1f with Fig.1g in
two momentum regions. One is the Glauber region:
\begin{equation}
  k^\mu \sim (\lambda^2, \lambda^2, \lambda, \lambda), \ \ \ \ \ k_g^\mu \sim (1 , \lambda^2, \lambda, \lambda),
\end{equation}
where the virtual gluon is a Glauber gluon. Another is the ultra-soft region:
\begin{equation}
  k^\mu \sim (\lambda^2, \lambda^2, \lambda^2, \lambda^2), \ \ \ \ \ k_g^\mu \sim (1 , \lambda^2, \lambda, \lambda),
\end{equation}
By calculating each numerator of each contribution we find that the
leading contribution comes only from the interference of Fig.1a with Fig.1g with
the virtual gluon as a Glauber gluon. It is at $\lambda^{-3}$.
All other contributions to $F_T^{(1)}$
are at orders higher than $\lambda^{-3}$. It is constructive to discuss here what will
be factorized into Sivers function. Because of the outgoing gluon is collinear
to the initial quark as indicated in Eq.(12), the Glauber region of the virtual gluon
in Fig.1a
corresponds the collinear region of the gluon emitted by the initial quark.
This gluon is collinear to the initial quark and has the final state interaction
with the outgoing quark through an exchange of a Glauber gluon. We will find
in our calculation of the Sivers function in Fig.2a, the emission of the collinear gluon
and the final state interaction is correctly contained in the Sivers function.
It is straightforward
to obtain the leading contribution:
\begin{eqnarray}
    F_{T}^{(1)}(x_B, z_h, P_{h\perp}) &=& -\frac{m\alpha_{s}^{2}(N_{C}^{2}-1)}{8\pi^{2}}
    \frac{\delta(1-z_{h})x_{B}(1-x_{B})}{\vert \vec {P}_{h\bot} \vert \left ( P_{h\bot}^{2}+m^{2}(1-x_{B})^{2} \right )}
    \ln\left ( \frac{m^{2}(1-x_{B})^{2}}{P_{h\bot}^{2}+m^{2}(1-x_{B})^{2}}\right ).
\end{eqnarray}
\par
To verify the factorization formula in Eq.(9) we consider Sivers function of the same target
as a quark with the momentum $P$ and the transverse polarization $s_\perp$. In order to generate
nonzero $k_\perp$ and an absorptive part, one needs exchange of two gluons at leading order. Nonzero
absorptive parts in the amplitude can be found by cutting diagrams. The contributions at leading order
of $\alpha_s$ are given by the interference of diagrams given in Fig.2.
\par
\begin{figure}[hbt]
\begin{center}
\includegraphics[width=14cm]{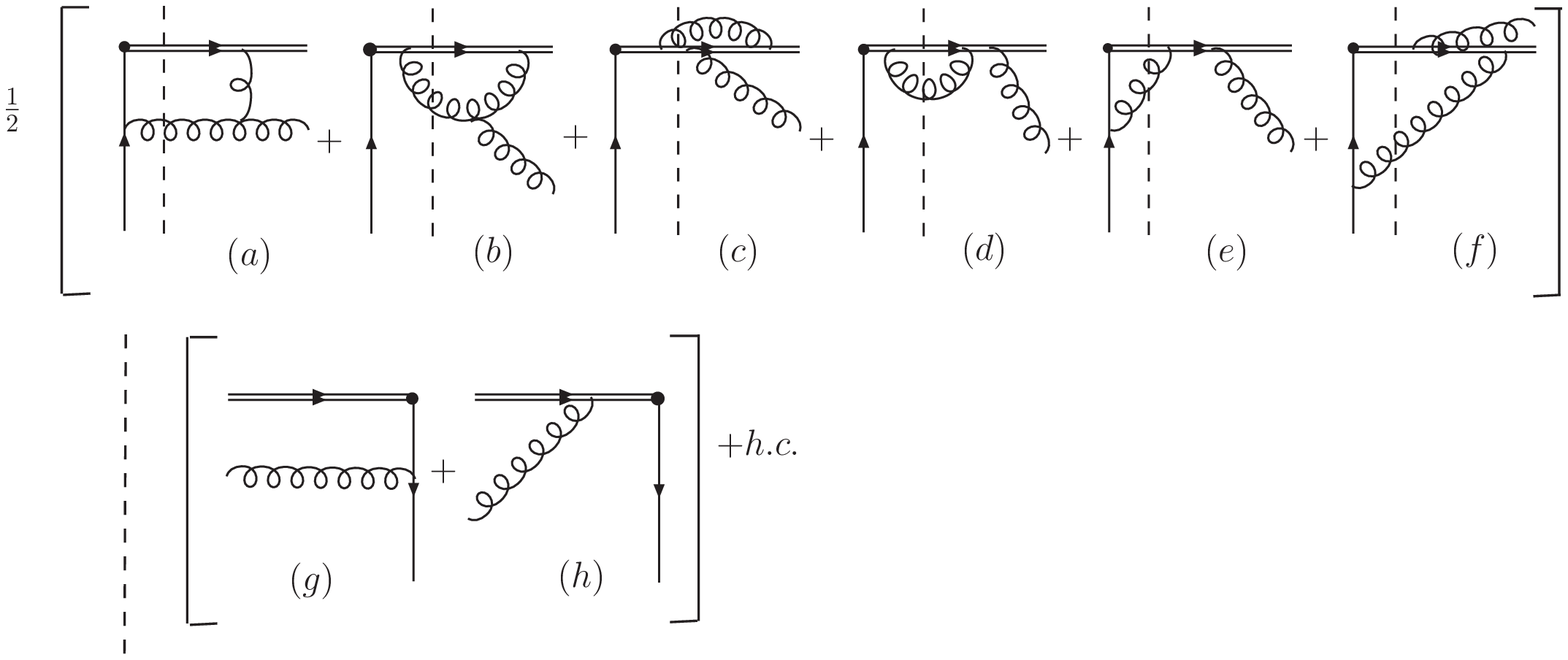}
\end{center}
\caption{Diagrams for the contributions to Sivers function. The broken lines
are cuts. The double lines represent the gauge links.  }
\label{Feynman-dg2}
\end{figure}
\par
In Fig.2 the diagrams in the first raw have an absorptive part indicated by the cut. It should be noted
that the gauge link here is pointing to the future representing a particle in the final state
with an infinity-large $-$-component of its momentum. Hence the energy flow is from left to right. This
is different than the case in Drell-Yan processes, where the gauge link represents
an incoming particle with an infinity-large $-$-component
of its momentum. This difference results in that the cut diagrams for Sivers function in SIDIS
are different than those for Sivers function appearing in Drell-Yan processes. This can be seen
by comparing Fig.2 here with Fig.1 and Fig.2 in \cite{MaSa}. Since we work at the leading order
of $\alpha_s$,
we can take the limit $u^+ \ll u^-$ directly here. After taking the limit, the Sivers function
receives the nonzero contribution only from the interference of Fig.2a with Fig.2g. This is
also in correspondence to the case of $F_T^{(1)}$. The detailed calculation can be done similarly
as done in \cite{MaSa}. We have:
\begin{eqnarray}
    q_{\bot}(x,k_{\bot})
     =-\frac{\alpha_{s}^{2}(N_{C}^{2}-1)}{8\pi^{2}}\frac{m x(1-x)}{k_{\bot}^{2}\left (k_{\bot}^{2}+m^{2}(1-x)^{2}
     \right )}
    \ln\left ( \frac{m^{2}(1-x)^{2}}{k_{\bot}^{2}+m^{2}(1-x)^{2}} \right ).
\end{eqnarray}
This result has been also obtained in \cite{SivF}.
Comparing the Sivers function for Drell-Yan processes in \cite{MaSa} we have the expected relation:
\begin{equation}
 q_\perp(x,k_{\bot})\vert_{SIDIS} = -q_{\bot}(x,k_{\bot})\vert_{DY},
\end{equation}
although it looks different from the corresponding cutting diagrams. The leading result of
$\hat q(x,p_\perp)$ can be found in \cite{JMY}:
\begin{equation}
 \hat q(x,p_\perp) = \delta (1-x) \delta^2(\vec p_\perp).
\end{equation}
With the partonic $F_T^{(1)}$ in Eq.(14) we verify the factorization formula for $F_T^{(1)}$
at the leading but nontrivial order of $\alpha_s$.
\par
Before turning to $F_T^{(2)}$, we briefly discuss the corresponding result for $F_T^{(1)}$ in
the collinear factorization. The formal derivation can be found in \cite{EKT,JQVY2}, where
the perturbative coefficient in the factorization is obtained. With our partonic result
for $F_T^{(1)}$ in Eq.(14) and that for the twist-3 matrix element calculated in \cite{MaSa},
we find that the collinear factorization for $F_T^{(1)}$ is the same as
for Drell-Yan processes given in \cite{MaSa}, but with a perturbative coefficient
different than that derived formally in \cite{JQVY2}. In this factorization
a twist-3 matrix element $T_F(x_1,x_2,\mu)$ defined in \cite{QiuSt,EFTE} appears.
With the result of
$T_F(x_1,x_2,\mu)$ defined in \cite{MaSa} we have the collinear factorization:
\begin{eqnarray}
F_T^{(1)}(x_B,z_h,P_{h\perp}) &=&
  \frac {\alpha_s N_c }{ 2 \pi^2 P_{h\perp}^3} \int \frac{dy_1}{y_1} \frac{ d y_2}{ y_2} D(y_2) \delta (1- \xi_2)
     \frac{ 1 }{(1-\xi_1)_+} x_B T_F(x_B,y_1,P_{h\perp}),
\end{eqnarray}
with $\xi_1 = x_B/y_1$, $\xi_2 =z_h/y_2$. In the above $D$ is the standard fragmentation function
in the collinear factorization. The same situation is also found in Drell-Yan processes\cite{MaSa}.
With the result presented in this work it is difficult to see why $F_T^{(1)}$
derived from our partonic results takes a different factorization form than that derived
with the diagram expansion at hadron level.
In the collinear factorization for SSA \cite{QiuSt, EFTE,KaKo,EKT}, there are three partons
entering hard scattering in cut diagrams for the differential cross section.
We have constructed a suitable state of massless partons in our later work\cite{CMS}
to study the difference, where there are also three partons entering hard scattering.
We have found that the so-called soft-pole contributions do not exist at the
leading order of $\alpha_s$ order and the gluon among the three partons is always
transversely polarized. This can be the reason for the difference. Details can be found
in \cite{CMS}.

\par\vskip20pt\noindent
{\bf 4.} Now we turn to $F_T^{(2)}$. In this case it becomes complicated because more than
one diagram can give nonzero contribution and also the loop integrals have a collinear divergence.
Since our purpose is to verify the factorization, we can first find the leading contributions
to $F_T^{(2)}$ from Fig.1 without performing the loop integrals, and then find
the corresponding contributions to Collins function to check the factorization.
The possible
contributions to Collins function are given by the diagrams in Fig.3.
In Fig.3 each of these diagrams is in one-to-one correspondence to each of those in Fig.1,
if one identifies the gauge link in Fig.3 as the initial quark line in Fig.1.
The result for $\delta q _\perp$ at leading order simply reads:
\begin{equation}
\delta q _T (x,k_\perp) = \delta(1-x) \delta^2 (\vec k_\perp).
\end{equation}
\par
\begin{figure}[hbt]
\begin{center}
\includegraphics[width=14cm]{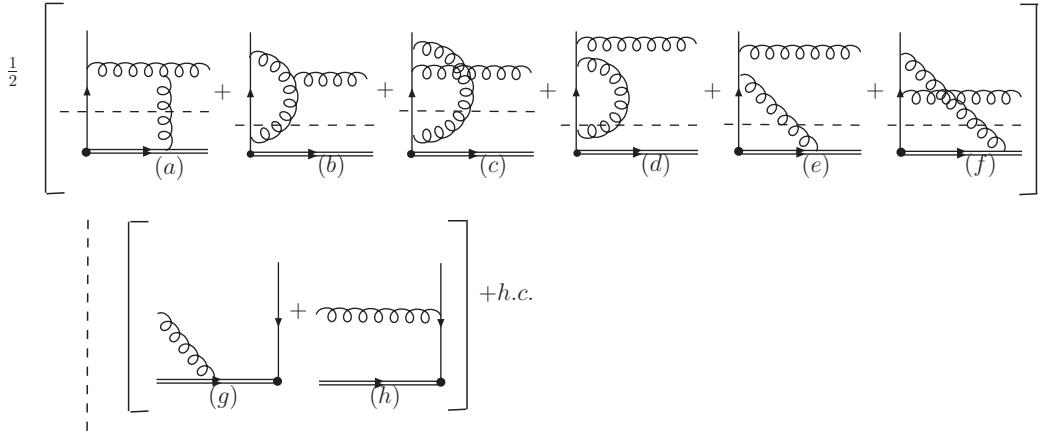}
\end{center}
\caption{Diagrams for the contributions to Collins function. The broken lines
are cuts. The double lines represent the gauge links.  }
\label{Feynman-dg3}
\end{figure}
\par
Let us look at the contribution from the interference of Fig.1a with Fig.1h. We denote
the momentum of the gluon crossing the cut in Fig.1a as $k$. In this contribution
the possible dominant contribution is with $k_g$ in the region collinear to
$P_h$, i.e., $k_g^\mu$ is scaled as $(\lambda^2, {\mathcal O}(1), \lambda,\lambda)$.
With the power counting
of the denominators and the integration measure of $k$ we find the possible dominant
contributions are from two regions of $k$. One is specified with $k$ as collinear
to $k_g$, another is with $k$ scaled as $(\lambda^2,\lambda^2,\lambda^2,\lambda^2)$.
Through calculating the corresponding trace in the numerator we find that
the leading contribution to $W^{\mu\nu}$ is only from the collinear region. We
extract the structure function $F_T^{(2)}$ as:
\begin{eqnarray}
F_T^{(2)} \vert_{ah} &=& -\frac{m\alpha_s^2}{4\pi} (N_c^2-1) \frac{ z_h \delta (1-x_B)}{\vert {\vec P}_{h\perp} \vert }
 \int \frac{ d^4 k}{(2\pi)^2} \delta (k^2) \delta ( (P_h +k_g -k)^2-m^2)
\nonumber\\
 && \cdot \frac{1}{P^-_{h} k^-}
   \frac{1}{(k_g-k)^2+i\varepsilon}
   \left [ k^- \left ( 2k_g^- -k^- +4 P_h^- \right ) -P^-_h \frac{\vec k_\perp \cdot {\vec k}_{g\perp}}{k^2_{g\perp}}
        \left ( 2k_g^- + k^- \right ) \right ] +\cdots,
\end{eqnarray}
where the $\cdots$ denote the terms suppressed by the power $\lambda$ or at higher orders of $m^2$.
The leading order of $F_T^{(2)} \vert_{ah}$ is at $\lambda^{-3}$ with $d=-4$ and $n=1$.
In the above integral one can use the two $\delta$-functions to perform
the integration of $k^+$ and $k^-$.
The integration over $k_\perp$ is bound from the above because of the momentum conservation.
We can extend the upper bound to infinity because we only need the leading
order contribution. We also note that the above integral has a collinear divergence when $k$ is collinear
to $k_g$.
The Collins function receives a contribution from the interference of Fig.3a with Fig.3h.
This contribution corresponds to the above contribution $F_T^{(2)} \vert_{ah}$.
Evaluating this contribution we have:
\begin{eqnarray}
\delta \hat q (z_h, P_{h\perp} )\vert_{ah} &=& -\frac{m\alpha_s^2}{4\pi} (N_c^2-1) \frac{ z_h }{P_{h\perp}^2 }
\int \frac{ d^4 k}{(2\pi)^2} \delta (k^2) \delta ( (P_h +k_g -k)^2-m^2)
\nonumber\\
 && \cdot \frac{1}{P_h^- k^-}
   \frac{1}{(k_g-k)^2+i\varepsilon}
   \left [ k^- \left ( 2k_g^- -k^- +4 P_h^- \right ) -P_h^- \frac{\vec k_\perp \cdot {\vec k}_{g\perp}}{k^2_{g\perp}}
        \left ( 2k_g^- + k^- \right ) \right ].
\end{eqnarray}
Comparing the above equations we find that the two contributions satisfy the factorization of $F_T^{(2)}$
in Eq.(9).
\par
Inspecting other contributions to $F_T^{(2)}$ in Fig.1, we find that the following contributions
are at higher order of $\lambda$: The interference of Fig.1g with Fig.1a, Fig.1d, Fig.1e and Fig.1f
and the interference of Fig.1e with Fig.1h. Calculating the correspond diagrams in Fig.3 in the limit
$v^+ \gg v^-$, we find
that the contributions to Collins function, corresponding to the above interferences, are zero.
Calculating the remaining diagrams in Fig.1 similar to the interference of Fig.1a with Fig.1h,
we find that the leading contributions can be factorized with the contributions
to Collins function in Fig.3 in an one-to-one correspondence.
Therefore, the factorization for
$F_T^{(2)}$ in Eq.(9) is verified with our partonic state, although we have not obtained detailed
results for $F_T^{(2)}$ and the Collins function with our parton states.
\par
In $e^+ e^-$-annihilations into hadrons, certain SSA is related to the Collins function which
is different than the Collins function in SIDIS. In $e^+ e^-$-annihilations the Collins function
is defined by replacing the past-pointing gauge links in Eq.(8) with the future-pointing gauge links.
It is interesting to ask if there is a relation between the two Collins function like the case
with Sivers functions in Eq.(16). The relation for Sivers functions in Eq.(16)
can be derived in general with Parity(P)- and Time(T)-reversal symmetry.
However, it seems impossible to obtain a relation between two Collins functions with PT symmetries, because
after PT-transformation the semi inclusive sum of out-states
becomes the semi-inclusive sum of in-states.
It is unclear how to relate a matrix element with such an in-state
to that of the corresponding out-state\cite{CAM}. In \cite{CAM} it is proposed that
one can use the future-pointing gauge links instead of the past-pointing gauge links
in fragmentation functions to perform the factorization. The proposal has been examined
at one-loop level and arguments beyond one-loop level have also been given\cite{CAM}.
This implies that the direction of the gauge links is irrelevant in the two Collins functions
and they are the same.
\par
Some model-calculations also have shown that the two Collins functions
are the same\cite{Metz,AG,FYuan}. In \cite{FYuan}
it has been shown that the different directions of the gauge links in the case
with one- and two gluon exchange make no difference in contributions to Collins functions.
From our calculation we can also conclude at the considered order that the Collins functions
are universal. The contributions in Fig.3 to the Collins function in SIDIS
are with the past-pointing gauge link. This gauge link represents a particle
with infinite $+$-energy moving from right to left in Fig.3.
If we change the gauge link into the future-pointing gauge link,
the future-pointing gauge link represents a particle with infinite $+$-energy moving from left to right.
With the future-pointing gauge link, the cut for Fig.3b, Fig.3c and Fig.3d
is the same and also there is no additional cut for these diagrams.
Hence the contributions from Fig.3b, Fig.3c and Fig.3d are the same for two Collins functions.
The contribution
of Fig.3e to Collins functions is always zero.
With the future-pointing gauge link, the two remaining diagrams, i.e., Fig.3a and Fig.3f
can have 3 cuts. One of them is the same as given in Fig.3.
This cut gives the same contribution.
The additional two cuts are cutting the gauge link and other propagator. It is interesting
that the contributions from the additional two cuts cancel each other. Similar cancelation
of contributions from additional cuts by changing the direction of gauge links
is also observed in a model-calculation\cite{Metz}.
Therefore, our result shows that
the two Collins functions are the same at the considered order.
\par\vskip20pt
\noindent
{\bf 5.} In this work we have verified the TMD factorization for SSA in SIDIS with partonic
states at the leading order of $\alpha_s$. Although it is at the leading order, but
it is nontrivial as shown in this work. Two structure functions related to SSA has been
examined. They take a factorized form in terms of TMD parton distributions and TMD parton fragmentation
functions. The factorized form agrees with that derived formally. However, as in the case of Drell-Yan processes,
we find that our partonic results satisfy a collinear factorization but with a perturbative coefficient
different than that derived formally. In this work and \cite{MaSa} we have made attempts to check
two factorization approaches with partonic states with a nonzero quark mass. It is worth
to point out that the same results of SSA related to Sivers function can also be derived
with massless partons as shown in our later work\cite{CMS}.

\par\vskip20pt
\par\noindent
{\bf\large Acknowledgments}
\par
We thank Prof. X.D.Ji, J.W. Qiu and F. Yuan for helpful discussions.
This work is supported by National Nature Science Foundation of P.R. China((No. 10721063,10575126).
\par\vskip30pt

\par

\end{document}